\documentclass[preprint,showpacs,superscriptaddress,nofootinbib]{revtex4}
\usepackage[dvips]{graphicx}
\usepackage{amsmath,amssymb,cancel}

\newcommand{\eV}{{\text{eV}}}
\newcommand{\keV}{{\text{keV}}}

\newcommand{\GeV}{{\text{GeV}}}
\newcommand{\TeV}{{\text{TeV}}}
\newcommand{\BR}{\text{BR}}
\newcommand{\tr}{{\text{tr}}}
\newcommand{\diag}{{\text{diag}}}

\newcommand{\U}{{\text{U}}}
\newcommand{\SU}{{\text{SU}}}
\newcommand{\MNS}{{\text{MNS}}}
\newcommand{\eff}{{\text{eff}}}
\newcommand{\nTHDM}{{$\nu$THDM}}

\begin{document}
\preprint{UT-HET-081}
\preprint{MISC-2013-05}

\title{
Loop Suppression of Dirac Neutrino Mass
in the Neutrinophilic Two Higgs Doublet Model
}

\author{Shinya Kanemura}
\email{kanemu@sci.u-toyama.ac.jp}
\affiliation{
Department of Physics,
University of Toyama, Toyama 930-8555, Japan
}
\author{Toshinori Matsui}
\email{matsui@jodo.sci.u-toyama.ac.jp}
\affiliation{
Department of Physics,
University of Toyama, Toyama 930-8555, Japan
}
\author{Hiroaki Sugiyama}
\email{sugiyama@cc.kyoto-su.ac.jp}
\affiliation{
Maskawa Institute for Science and Culture,
Kyoto Sangyo University, Kyoto 603-8555, Japan
}


\begin{abstract}
 We extend the scalar sector of
the neutrinophilic two Higgs doublet model,
where small masses of Dirac neutrinos
are obtained via a small vacuum expectation value $v_\nu$
of the neutrinophilic $\SU(2)_L$-doublet scalar field
which has a Yukawa interaction with only right-handed neutrinos.
 A global $\U(1)_X$ symmetry is used
for the neutrinophilic nature of
the second $\SU(2)_L$-doublet scalar field
and also for eliminating Majorana mass terms of neutrinos.
 By virtue of an appropriate assignment
of the $\U(1)_X$-charges to new particles,
our model has an unbroken $Z_2$ symmetry,
under which the lightest $Z_2$-odd scalar boson
can be a dark matter candidate.
 In our model,
$v_\nu$ is generated by the one-loop diagram
to which $Z_2$-odd particles contribute.
 We briefly discuss a possible signature of our model at the LHC\@.
\end{abstract}

%

\maketitle

\section{Introduction}
\label{Sec:intro}

 It has been well established
that neutrinos have nonzero masses
as shown in the neutrino oscillation measurements%
~\cite{Ref:solar-v, Ref:atom-v, Ref:acc-v, Ref:acc-app-v,
Ref:short-reac-v, Ref:long-reac-v}
although they are massless particles
in the standard model~(SM) of particle physics.
 Since the scale of neutrino masses is
much different from that of the other fermion masses,
they might be generated
by a different mechanism
from the one for the other fermions.
 Usually,
the possibility that neutrinos are Majorana fermions
is utilized as a characteristic feature
of the neutrino masses.
 The most popular example is the seesaw mechanism~\cite{Ref:Type-I}
where very heavy right-handed Majorana neutrinos are introduced.
 However,
lepton number violation
which is caused by masses of the Majorana neutrinos
has not been discovered.
 Thus
it is worth considering
the possibility that neutrinos are not Majorana fermions
but Dirac fermions similarly to charged fermions.

 The neutrinophilic two Higgs doublet model~(\nTHDM)
is a new physics model where neutrinos are regarded as
Dirac fermions.
 The second $\SU(2)_L$-doublet scalar field
which couples only with right-handed neutrinos $\nu_R$
was first introduced in Ref.~\cite{Ma:2000cc}
for Majorana neutrinos.
 Phenomenology in the model of Majorana neutrinos
is discussed in Ref.~\cite{Ref:nuTHDM-M,Haba:2011fn}.
 The neutrinophilic doublet field
is also utilized for Dirac neutrinos~\cite{Ref:nuTHDM-D}
where a spontaneously broken $Z_2$ parity
is introduced in order to achieve the neutrinophilic property.
 Smallness of neutrino masses
are explained by a tiny vacuum expectation value~(VEV)
of the neutrinophilic scalar
without extremely small Yukawa coupling constant for neutrinos.
 Instead of the $Z_2$ parity,
the model in Ref.~\cite{Davidson:2009ha} uses
a global $\U(1)_X$ symmetry that is softly broken
in the scalar potential.
 The $\U(1)_X$ symmetry forbids Majorana mass terms of $\nu_R$,
and then neutrinos are Dirac fermions%
\footnote{
 Since the Majorana mass terms of $\nu_R$
can also be acceptable as soft breaking terms of the $\U(1)_X$,
the lepton number conservation may be imposed to the Lagrangian.
}.
 We refer to the model in Ref.~\cite{Davidson:2009ha} as the \nTHDM\@.

 The new particle which was discovered
at the LHC~\cite{Aad:2012tfa,Chatrchyan:2012ufa}
is likely to be the SM Higgs boson%
~\cite{ATLAS:2013sla,ATLAS:2013mla,CMS:aya,Chatrchyan:2012jja}.
 It opens the new era of probing
the origin of particle masses.
 Then
it would be a natural desire to expect that
the origin of neutrino masses are also uncovered.
 If the neutrinophilic scalars in the \nTHDM\ exist
within the experimentally accessible energy scale~(namely the TeV-scale),
decays of the neutrinophilic charged scalar into leptons
can provide direct information on the neutrino mass matrix
because it is proportional to
the matrix of new Yukawa coupling constants
for the neutrinophilic scalar field~\cite{Davidson:2009ha,Davidson:2010sf}.
 In such a case,
the smallness of a new VEV
which is relevant to Dirac neutrino masses
is interpreted by the smallness of
a soft-breaking parameter of the global $\U(1)_X$ symmetry.
 It seems then better to have a suppression mechanism
for the soft-breaking parameter
by extending the \nTHDM\ with TeV scale particles
including a dark matter candidate.
 The existence of dark matter has also been established
in cosmological observations~\cite{Hinshaw:2012aka,Ade:2013zuv},
and it is an important guideline
for constructing new physics models.

 The reason why the neutrino masses are tiny
can be explained by a mechanism
that the interaction of neutrinos with the SM Higgs boson
is generated via a loop diagram involving a dark matter candidate in the loop
while the interaction is forbidden at the tree level%
~\cite{KNT, Ref:Ma, Ma:2007gq, Ma:2008cu, AKX, Gustafsson:2012vj, Aoki:2011yk,
Ref:highD-op, Gu:2007ug, Ref:rad-seesaw, Kajiyama:2013zla}.
 Notice that
smallness of neutrino masses
in such radiative mechanisms
does not require new particles to be very heavy.
 Similarly,
if neutrino masses arise from a new VEV,
smallness of neutrino masses
can be explained by assuming that
the VEV is generated at the loop level
by utilizing a dark matter candidate~\cite{Kanemura:2012rj}.
 In this paper,
we extend the \nTHDM\
such that the new VEV is generated at the one-loop level
(see also Ref.~\cite{Chang:1986bp})
where a dark matter candidate is involved in the loop.

 This paper is organized as follows.
 We briefly introduce the \nTHDM\ in Sec.~\ref{Sec:nuTHDM}.
 The \nTHDM\ is extended in Sec.~\ref{Sec:1-loop}
such that a small VEV is generated via the one-loop diagram
which involving a dark matter candidate in the loop.
 Section~\ref{Sec:pheno}
is devoted to discussion on phenomenology
in the extended \nTHDM\@.
 We conclude in Sec.~\ref{Sec:concl}.

\section{Neutrinophilic Two-Higgs-Doublet Model}
\label{Sec:nuTHDM}

 In the \nTHDM,
the SM is extended with
the second $\SU(2)_L$-doublet scalar field $\Phi_\nu$
which has a hypercharge $Y=1/2$
and right-handed neutrinos $\nu_{i R}^{}$~($i = 1\text{-}3$)
which are singlet fields under the SM gauge group.
 A global $\U(1)_X$ symmetry is introduced,
under which $\Phi_\nu$ and $\nu_{i R}^{}$ have the same nonzero charge
while the SM particles have no charge.
 Then,
the Yukawa interaction with $\Phi_\nu$
is only the following one:
\begin{eqnarray}
{\mathcal L}_\text{$\nu$-Yukawa}
=
 - (y_\nu^{})_{\ell i}
 \overline{L_\ell}\, i\sigma_2\, \Phi_\nu^\ast\, \nu_{iR}^{}
 + \text{h.c.} ,
\label{Eq:nuYukawa}
\end{eqnarray}
where $\ell (=e, \mu, \tau)$ denotes the lepton flavor
and $\sigma_i~(i=1\text{-}3)$ are the Pauli matrices.
 Since Majorana mass terms $\overline{(\nu_{iR}^{})^c}\,\nu_{iR}$
are forbidden by the $\U(1)_X$ symmetry,
there appears an accidental conservation of the lepton number
where lepton numbers of $\Phi_\nu$ and $\nu_{iR}^{}$
are $0$ and $1$, respectively.
 When the neutral component $\phi_\nu^0$ of $\Phi_\nu$
develops its VEV
$v_\nu$~($\equiv \sqrt{2}\, \langle \phi_\nu^0 \rangle$),
the neutrino mass matrix arise as
$(m_\nu)_{\ell i} = v_\nu (y_\nu^{})_{\ell i}/\sqrt{2}$.
 We have taken a basis
where $\nu_{iR}^{}$ are mass eigenstates.
 Then
the mass matrix $m_\nu^{}$ is diagonlized as
$U_\MNS^\dagger m_\nu^{} = \diag(m_1, m_2, m_3)$,
where $m_i~(i=1\text{-}3)$ are the neutrino mass eigenvalues
and a unitary matrix $U_\MNS$
is the so-called Maki-Nakagawa-Sakata~(MNS) matrix~\cite{Maki:1962mu}.
 Dirac neutrinos are constructed as
$\nu_i^{}
= (\sum_\ell (U_\MNS^\dagger)_{i\ell} \nu_{\ell L}^{} , \ \nu_{iR}^{})^T$.
 Smallness of neutrino masses
is attributed to that
$v_\nu$ is much smaller than $v$.

 If the VEV $v_\nu$ is generated spontaneously,
a CP-odd scalar $\phi_{\nu i}^0$ becomes massless
as a Nambu-Goldstone boson
with respect to the breaking of $\U(1)_X$,
where $\phi_\nu^0 = (v_\nu + \phi_{\nu r}^0 + i \phi_{\nu i}^0)/\sqrt{2}$.
 In addition,
a CP-even neutral scalar $\phi_{\nu r}^0$
has a small mass ($\propto v_\nu \ll v$).
 Therefore,
the scenario of the spontaneous breaking of $\U(1)_X$
is not allowed by the measurement
of the invisible decay of the $Z$ boson.
 The scalar potential in the \nTHDM\ is given by
\begin{eqnarray}
V^{\text{(\nTHDM)}}
&=&
 - \mu_{\Phi 1}^2 \Phi^\dagger \Phi
 + \mu_{\Phi 2}^2 \Phi_\nu^\dagger \Phi_\nu
 - \left(
    \mu_{\Phi 12}^2 \Phi_\nu^\dagger \Phi
    + \text{h.c.}
   \right)
\nonumber\\
&&{}
 + \lambda_{\Phi 1} (\Phi^\dagger \Phi)^2
 + \lambda_{\Phi 2} (\Phi_\nu^\dagger \Phi_\nu)^2
 + \lambda_{\Phi 12} (\Phi^\dagger \Phi)(\Phi_\nu^\dagger \Phi_\nu)
 + \lambda_{\Phi 12}^\prime (\Phi^\dagger \Phi_\nu)(\Phi_\nu^\dagger \Phi) ,
\end{eqnarray}
where $\mu_{\Phi 12}^2$ can be real and positive
by using rephasing of $\Phi_\nu$
without loss of generality;
 We take $\mu_{\Phi 1}^2 > 0$ and $\mu_{\Phi 2}^2 > 0$.
 The VEV of $\phi_\nu^0$ is triggered
by $\mu_{\Phi 12}^2$ which softly breaks the $\U(1)_X$ symmetry.
 Since the term does not breaks
the lepton number conservation,
neutrinos are still Dirac particles.
 Taking $v_\nu/v \ll 1$ into account,
the VEVs are calculated as
\begin{eqnarray}
v
\simeq
 \frac{\mu_{\Phi 1}^{}}{\sqrt{\lambda_{\Phi 1}}} , \quad
v_\nu^{}
\simeq
 \frac{\displaystyle
       2 v\, \mu_{\Phi 12}^2
      }
      {\displaystyle
       2 \mu_{\Phi 2}^2
       + (\lambda_{\Phi 12} + \lambda_{\Phi 12}^\prime) v^2
      } .
\end{eqnarray}
 If $\mu_{\Phi 2}^{} \sim v$,
we have $v_\nu^{} \sim \mu_{\Phi 12}^2/v$.
Then,
$\mu_{\Phi 12}^{}/v$ is required to be small
($\sim 10^{-6}$ for $y_\nu \sim 1$).
 Stability of the tiny $v_\nu^{}$
is discussed in Refs.~\cite{Haba:2011fn, Morozumi:2011zu}.
 In our model presented in the next section,
$\mu_{\Phi 12}^{}/v$ becomes small
because $\mu_{\Phi 12}^2$ is generated at the one-loop level.

\section{An Extension of the \nTHDM}
\label{Sec:1-loop}

 Since we try to generate $\mu_{\Phi 12}^2$ at the loop level,
it does not appear in the Lagrangian.
 Then
the $\U(1)_X$ symmetry should be broken spontaneously.
 For the spontaneous breaking,
we rely on an additional scalar $s_1^0$
which is a singlet field under the SM gauge group.
 Similarly to the singlet Majoron model~\cite{Ref:majoron}
where a VEV of a singlet field
spontaneously breaks the lepton number conservation
by two units,
the Nambu-Goldstone boson from $s_1^0$ is acceptable~\cite{Ref:majoron};
 the Nambu-Goldstone boson
couples first with only neutrinos among fermions.
 If $\U(1)_X$-charges of $\Phi_\nu$ and $s_1^0$ are
$3$ and $1$, respectively,
a dimension-5 operator $(s_1^0)^3 \Phi_\nu^\dagger \Phi$
is allowed by the $\U(1)_X$ symmetry
although $\Phi_\nu^\dagger \Phi$ is forbidden.
 Then,
$\mu_{\Phi 12}^2$ is generated from the dimension-5 operator
with the VEV of $s_1^0$.
 In this paper,
we show the simplest realization
of the dimension-5 operator at the one-loop level
where dark matter candidates are involved in the loop.

\begin{table}[t]
\begin{center}
\begin{tabular}{c||c|c||c|c|c}
 {}
 & \ $\nu_{iR}^{}$ \
 &
 \ $\Phi_\nu
    = \begin{pmatrix}
       \phi_\nu^+\\
       \phi_\nu^0
      \end{pmatrix}
   $ \
 &
 \ $\eta
    = \begin{pmatrix}
       \eta^+\\
       \eta^0
      \end{pmatrix}
   $ \
 & \ $s_1^0$ \
 & \ $s_2^0$ \
\\\hline\hline
 $SU(2)_L$
 & {\bf \underline{1}}
 & {\bf \underline{2}}
 & {\bf \underline{2}}
 & {\bf \underline{1}}
 & {\bf \underline{1}}
\\\hline
 $U(1)_Y$
 & $0$
 & $1/2$
 & $1/2$
 & $0$
 & $0$
\\\hline\hline
 Global $U(1)_X$ \
 & \ $3$ \
 & \ $3$ \
 & \ $3/2$ \
 & \ $1$ \
 & \ $1/2$\ 
\end{tabular}
\end{center}
\caption{
 New particles which are added to the SM
in our model.
}
\label{Tab:particles}
\end{table}

 Table~\ref{Tab:particles}
is the list of new particles
added to the SM\@.
 In the table,
$\nu_{iR}^{}$ and $\Phi_\nu$
are the particles which exist in the \nTHDM\@.
 The $\U(1)_X$ symmetry is spontaneously broken
by the VEV of $s_1^0$.
 We take a scenario where
$\eta$ and $s_2^0$ do not have VEVs.
 Since their $\U(1)_X$-charges are half-integers
while the one for $s_1^0$ is an integer,
a $Z_2$ symmetry remains unbroken after the $\U(1)_X$ breaking.
 Here, $\eta$ and $s_2^0$ are $Z_2$-odd particles.
 The $Z_2$ symmetry stabilizes
the lightest $Z_2$-odd particle
which can be a dark matter candidate.

 The Yukawa interaction
in this model is identical to
those in the \nTHDM~(see Eq.~\eqref{Eq:nuYukawa}).
 The scalar potential in this model is expressed as
\begin{eqnarray}
V
&=&
 - \mu_{s1}^2 | s_1^0 |^2
 + \mu_{s2}^2 | s_2^0 |^2
 - \mu_{\Phi 1}^2 \Phi^\dagger \Phi
 + \mu_{\Phi 2}^2 \Phi_\nu^\dagger \Phi_\nu
 + \mu_\eta^2 \eta^\dagger \eta
\nonumber\\
&&{}
 - \left(
    \mu\, s_1^{0\ast} (s_2^0)^2 + \text{h.c.}
   \right)
\nonumber\\
&&{}
 + \left(
    \lambda_{s \Phi1 \eta}\,
    s_1^{0\ast} (s_2^0)^\ast \Phi^\dagger \eta
    + \text{h.c.}
   \right)
 + \left(
    \lambda_{s \Phi2 \eta}\,
    s_1^0 s_2^0 \Phi_\nu^\dagger \eta
    + \text{h.c.}
   \right)
 + \cdots .
\label{Eq:V-1}
\end{eqnarray}
 Only the relevant parts to our discussion are presented in Eq.~\eqref{Eq:V-1}.
 The other terms are shown in Appendix.
 Parameters $\mu$, $\lambda_{s \Phi1 \eta}$, and $\lambda_{s \Phi2 \eta}$
are taken to be real and positive values
by rephasing of scalar fields
without loss of generality.
 At the tree level,
$v_\nu$, $v$, and $v_s~(= \sqrt{2} \langle s_1^0 \rangle)$
are given by
\begin{eqnarray}
v_\nu = 0 , \quad
\begin{pmatrix}
 v^2\\
 v_s^2
\end{pmatrix}
=
 \frac{2}{ 4 \lambda_{s1} \lambda_{\Phi 1} - \lambda_{s1\Phi 1}^2 }
 \begin{pmatrix}
  2\lambda_{s1} & -\lambda_{s1\Phi 1}\\[1mm]
  -\lambda_{s1\Phi 1} & 2\lambda_{\Phi 1}
 \end{pmatrix}
 \begin{pmatrix}
  \mu_{\Phi 1}^2\\[1mm]
  \mu_{s1}^2
 \end{pmatrix} .
\end{eqnarray}

 The $Z_2$-odd scalar fields ($\eta$ and $s_2^0$)
result in the following particles:
two CP-even neutral scalars
(${\mathcal H}_1^0$ and ${\mathcal H}_2^0$),
two CP-odd neutral ones (${\mathcal A}_1^0$ and ${\mathcal A}_2^0$),
and a pair of charged ones (${\mathcal H}^\pm$).
 It is clear that ${\mathcal H}^\pm = \eta^\pm$.
 When ${\mathcal H}_1^0$ (or ${\mathcal A}_1^0$)
is lighter than ${\mathcal H}^\pm$,
the neutral one becomes the dark matter candidate.
 On the other hand,
from $Z_2$-even scalar fields ($\Phi$, $\Phi_\nu$, and $s_1^0$),
we have three CP-even particles ($h^0$, $H^0$, and $H_\nu^0$),
two CP-odd ones ($A_\nu^0$ and a massless $z_2^0$),
and a pair of charged scalars ($H_\nu^\pm$).
 The mixings between $\phi_\nu^0$ and others are ignored
because we take $v_\nu/v \ll 1$ and $v_\nu/v_s \ll 1$.
 Then,
$\Phi_\nu$ provides $H_\nu^0~(=\phi_{\nu r}^0)$,
$A_\nu^0~(=\phi_{\nu i}^0)$,
and $H_\nu^\pm~(=\phi_\nu^\pm)$.
 It is easy to see that $z_2^0 = s_{1i}^0$,
where
$s_1^0
=
 \left(
  v_s^{} + s_{1r}^0 + i s_{1i}^0
 \right)/\sqrt{2}
$.
 The formulae of scalar mixings and scalar masses
are presented in Appendix.
 Hereafter,
we assume that scalar fields in Tab.~\ref{Tab:particles}
are almost mass eigenstates just for simplicity,
which is achieved
when $\lambda_{s\Phi 1\eta}$ and $\lambda_{s1\Phi 1}$ are small.

\begin{figure}[t]
\begin{center}
\includegraphics[scale=0.8]{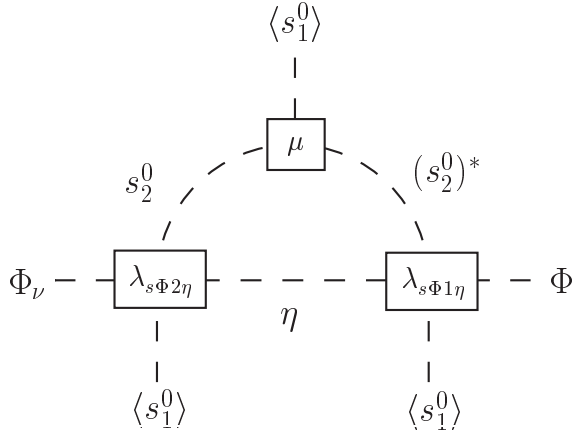}
\vspace*{-4mm}
\caption{
 The one-loop diagram of
the leading contribution to
$(\mu_{\Phi 12}^2)_\eff\, [ \Phi_\nu^\dagger \Phi ]$
with respect to $\mu$, $\lambda_{s\Phi 1\eta}^{}$,
and $\lambda_{s\Phi 2\eta}^{}$.
}
\label{Fig:loop}
\end{center}
\end{figure}

 By using cubic and quartic interactions
shown in Eq.~\eqref{Eq:V-1},
the interaction $\Phi_\nu^\dagger \Phi$
is obtained with the one-loop diagram in Fig.~\ref{Fig:loop}.
 The coefficient $(\mu_{\Phi 12}^2)_\eff$ of the interaction
is calculated as
\begin{eqnarray}
(\mu_{\Phi 12}^2)_\eff
=
 \frac{
       \mu\, \lambda_{s \Phi 1 \eta}\, \lambda_{s \Phi 2 \eta}\,
       v_s^3
      }
      { 32\sqrt{2}\, \pi^2 (m_\eta^2 - m_{s2}^2) }
 \left(
  1
  - \frac{ m_\eta^2 }{ m_\eta^2 - m_{s2}^2 }
    \ln\frac{m_\eta^2}{m_{s2}^2}
 \right) ,
\end{eqnarray}
where
\begin{eqnarray}
m_\eta^2
&\equiv&
 \mu_\eta^2
 + \frac{1}{\,2\,}
   \Bigl\{
    \left(
     \lambda_{\Phi 1\eta} + \lambda_{\Phi 1\eta}^\prime
    \right) v^2
    + \lambda_{s1\eta} v_s^2
   \Bigr\} ,
\\
m_{s2}^2
&\equiv&
 \mu_{s2}^2
 + \frac{1}{\,2\,}
   \left(
    \lambda_{s2\Phi 1} v^2
    + \lambda_{s12} v_s^2
   \right) .
\end{eqnarray}
 Ignoring loop corrections
to terms which exist at the tree-level,
we finally arrive at
\begin{eqnarray}
v_\nu^{}
=
 \frac{ v\, (\mu_{\Phi 12}^2)_\eff }{ m_{H_\nu^0}^2 } ,
\end{eqnarray}
where
$m_{H_\nu^0}^2
\equiv
 \mu_{\Phi 2}^2
 + \frac{1}{2} (\lambda_{\Phi 12} + \lambda_{\Phi 12}^\prime) v^2
 + \frac{1}{2} \lambda_{s1 \Phi 2} v_s^2$
which is the mass of $H_\nu^0~(=\phi_{\nu r}^0)$.
 For example,
we have $m_\nu = {\mathcal O}(0.1)\,\eV$
for $m_{s2}^{} = {\mathcal O}(10)\,\GeV$ (as the dark matter mass),
$v_s^{} \sim m_\eta^{} \sim m_{H_\nu}^{} = {\mathcal O}(100)\,\GeV$,
$\mu = {\mathcal O}(1)\,\GeV$,
$y_\nu = {\mathcal O}(10^{-4})$,
and $\lambda_{s\Phi 1\eta}^{} \sim \lambda_{s\Phi 2 \eta}^{}
= {\mathcal O}(10^{-2})$.

\section{Phenomenology}
\label{Sec:pheno}

 Hereafter,
we take the following values of parameters
as an example:
\begin{equation}
\begin{split}
&
(y_\nu)_{\ell i}
\sim 10^{-4} , \quad
\lambda_{s\Phi 1\eta}^{} = \lambda_{s\Phi 2 \eta}^{}
= 10^{-2} , \quad
\mu
= 1\,\GeV , \quad
v_s
= 300\,\GeV , \\
&
m_{H_\nu^0}^{} = m_{A_\nu^0}^{} = m_{H_\nu^\pm}^{}
= 300\,\GeV , \quad
m_{{\mathcal H}_2^0}^{}
= 230\,\GeV , \quad
m_{{\mathcal H}_1^0}^{} = 60\,\GeV .
\end{split}
\label{Eq:benchmark}
\end{equation}
 These values can satisfy constraints from
the $\rho$ parameter,
searches of lepton flavor violating processes,
the relic abundance of dark matter,
and direct searches for dark matter.
 In order to satisfy $\rho \simeq 1$,
particles which come from an $\SU(2)$ multiplet
have a common mass.
 If ${\mathcal H}_1^0 \simeq \eta_r^0$ for example,
we take
$m_{{\mathcal H}^\pm}^{} \sim
 m_{{\mathcal A}_1^0}^{} \sim
 m_{{\mathcal H}_1^0}^{}$.
 Since $y_\nu$ is not assumed to be very large,
contributions of $H_\nu^\pm$ to lepton flavor violating decays
of charged leptons are negligible.
 For example,
the branching ratio $\BR(\mu \to e\gamma)$~\cite{Davidson:2009ha}
is proportional to $|(y_\nu^{} y_\nu^\dagger)_{\mu e}|^2$
and becomes about $10^{-22}$
which is much smaller than the current bound
at the MEG experiment~\cite{Adam:2013mnn}:
$\BR(\mu \to e\gamma) < 5.7\times 10^{-13}$
at the $90\,\%$ confidence level.

\subsection{Dark Matter}
\label{Subsec:DM}

 We assume that
the mixing between $s_2^0$ and $\eta^0$ is negligible
for simplicity,
which corresponds to the case $\lambda_{s\Phi 1 \eta}^{} \ll 1$.
 Then,
the dark matter candidate ${\mathcal H}_1^0$
is dominantly made from $s_{2r}^0$ or $\eta_r^0$.
 We also assume that $\lambda_{s12} | s_1^0 |^2 | s_2^0 |^2$
and $\lambda_{s1\eta} | s_1^0 |^2 (\eta^\dagger \eta)$
are negligible
in order to avoid
${\mathcal H}_1^0 {\mathcal H}_1^0 \to z_2^0 z_2^0$
which would reduce the dark matter abundance too much.
 Notice that these coupling constants
($\lambda_{s12}$ and $\lambda_{s1\eta}$)
are not used in the loop diagram in Fig.~\ref{Fig:loop}.
 When ${\mathcal H}_1^0 \simeq s_{2r}^0$,
the ${\mathcal H}_1^0$ is similar to
the real singlet dark matter in Ref.~\cite{Silveira:1985rk}.
 Experimental constraints on the singlet dark matter
can be found e.g.\ in Ref.~\cite{Ref:const-ISM}.
 We see that
$53\,\GeV \lesssim m_{{\mathcal H}_1^0}^{} \lesssim 64\,\GeV$
and $90\,\GeV \lesssim m_{{\mathcal H}_1^0}^{}$
are allowed.
 On the other hand,
when ${\mathcal H}_1^0 \simeq \eta_r^0$,
the dark matter is similar to the one
in the so-called inert doublet model%
~\cite{Deshpande:1977rw,Barbieri:2006dq}.
 See e.g.\ Refs.~\cite{Ref:const-IDM,Gustafsson:2010zz}
for experimental constraints on the inert doublet model.
 It is shown that
$45\,\GeV \lesssim m_{{\mathcal H}_1^0}^{} \lesssim 80\,\GeV$ is allowed.
 In order to suppress
the scattering of ${\mathcal H}_1^0$ on nuclei
mediated by the $Z$ boson,
sufficient splitting of
$m_{{\mathcal H}_1^0}^{}$ and $m_{{\mathcal A}_1^0}^{}$
is required:
$m_{{\mathcal A}_1^0}^{} - m_{{\mathcal H}_1^0}^{} \gtrsim 100\,\keV$
(See e.g.\ Ref.~\cite{Gustafsson:2010zz}).
 Values of $m_{{\mathcal H}_1^0}^{}$ and $m_{{\mathcal H}_2^0}^{}$
in Eq.~\eqref{Eq:benchmark} are obtained
by using $m_\eta^{} = 60\,\GeV$ and $m_s^{} = 231\,\GeV$
in Eqs.~\eqref{Eq:mH1} and \eqref{Eq:mH2} in Appendix,
and then these values of $m_\eta^{}$ and $m_s^{}$
give $m_{{\mathcal A}_1^0}^{} - m_{{\mathcal H}_1^0}^{} \simeq 400\,\keV$.

 Since we discuss in the next subsection
a possible collider signature
where $H_\nu^0$ decays into ${\mathcal H}_1^0$,
a light dark matter ($m_{{\mathcal H}_1^0}^{} \simeq m_{h^0}^{}/2$)
is interesting
such that $H_\nu^0$ (and $H_\nu^\pm$) can also be light.
 We take $m_{{\mathcal H}_1^0}^{} = 60\,\GeV$ as an example for both cases,
${\mathcal H}_1^0 \simeq s_{2r}^0$
and ${\mathcal H}_1^0 \simeq \eta_r^0$.

\subsection{Collider}
\label{Subsec:col}

\begin{figure}
\begin{center}
\includegraphics[scale=0.8]{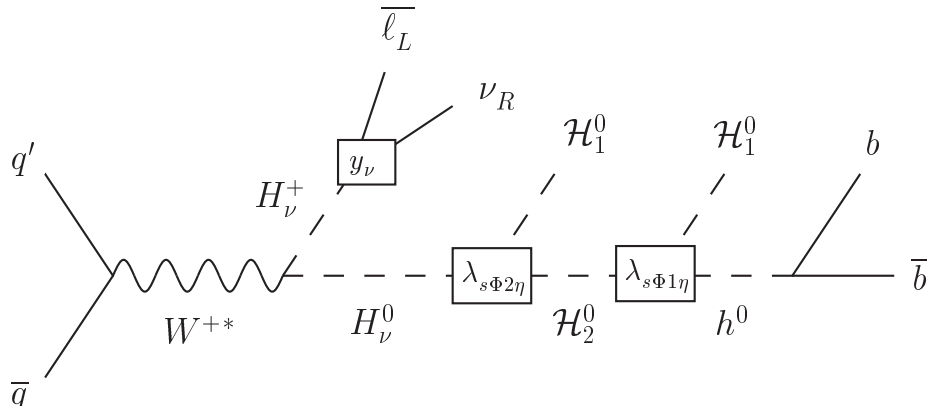}{}
\vspace*{-4mm}
\caption{
 A possible signature of our model at the LHC\@.
}
\label{Fig:LHC}
\end{center}
\end{figure}

 In the \nTHDM\ as well as in our model,
the neutrino mass matrix $m_\nu^{}$
is simply proportional to $y_\nu^{}$.
 The flavor structure of $H_\nu^+ \to \overline{\ell_L} \nu_R$
(summed over the neutrinos)
is predicted~\cite{Davidson:2009ha}
by using current information on $m_\nu^{}$
obtained by neutrino oscillation measurements.
 The prediction enables the \nTHDM\
to be tested at collider experiments.
 Since this advantage should not be spoiled,
$H_\nu^\pm \to {\mathcal H}_1^0 {\mathcal H}^\pm$
(${\mathcal H}^\pm {\mathcal H}_2^0$)
should be forbidden
for ${\mathcal H}_1^0 \simeq s_{2r}^0$ (${\mathcal H}_1^0 \simeq \eta_r^0$).
 Therefore, we assume that
$m_{{\mathcal H}^\pm}^{}$ satisfies
$m_{H_\nu^\pm}^{} \leq m_{{\mathcal H}_1^0}^{} + m_{{\mathcal H}^\pm}^{}$
for ${\mathcal H}_1^0 \simeq s_{2r}^0$
or $m_{H_\nu^\pm}^{} \leq m_{{\mathcal H}^\pm}^{} + m_{{\mathcal H}_2^0}^{}$
for ${\mathcal H}_1^0 \simeq \eta_r^0$;
 for example,
$m_{{\mathcal H}^\pm}^{} = 250\,\GeV$~(100\,\GeV)
for ${\mathcal H}_1^0 \simeq s_{2r}^0$~($\eta_r^0$).

 The process in Fig.~\ref{Fig:LHC}
would be a characteristic collider signature of our model.
 Notice that
the process utilizes
two coupling constants
($\lambda_{s\Phi 1\eta}^{}$ and $\lambda_{s\Phi 2\eta}^{}$)
which appear also in Fig.~\ref{Fig:loop}.
 Thus,
the process indicates that
$\mu_{\Phi 12}^2 \Phi_\nu^\dagger \Phi$
is radiatively generated with
a contribution of dark matter.
 In the original \nTHDM\ in comparison,
$H_\nu^0$ decays into $\nu \overline{\nu}$
for the case with $m_{H_\nu^0}^{} = m_{H_\nu^\pm}^{}$.
 In order to observe the process in Fig.~\ref{Fig:LHC},
the partial decay width
$\Gamma( H_\nu^0 \to {\mathcal H}_1^0 {\mathcal H}_2^0 )$
should be larger than
$\Gamma( H_\nu^0 \to \nu \overline{\nu} )$.
 Using our benchmark values,
we have
\begin{eqnarray}
\Gamma( H_\nu^0 \to \nu \overline{\nu} )
&=&
 \frac{ \tr(y_\nu^\dagger y_\nu^{}) m_{H_\nu^0}^{} }{ 16 \pi }
\ \simeq \ 60\,\eV ,
\\
\Gamma( H_\nu^0 \to {\mathcal H}_1^0 {\mathcal H}_2^0 )
&=&
 \frac{ \lambda_{s\Phi 2\eta}^2 v_s^2 }{ 64 \pi m_{H_\nu^0}^{} }
 \sqrt{
       1
       - \frac{ (m_{{\mathcal H}_2^0}^{}+m_{{\mathcal H}_1^0}^{})^2 }
              { m_{H_\nu^0}^2 }
      }
 \sqrt{
       1
       - \frac{ (m_{{\mathcal H}_2^0}^{}-m_{{\mathcal H}_1^0}^{})^2 }
              { m_{H_\nu^0}^2 }
      }
\ \simeq \
30\,\keV .
\end{eqnarray}
 Then,
$H_\nu^0$ decays into ${\mathcal H}_1^0 {\mathcal H}_2^0$ dominantly%
\footnote{
 Cascade decay of $A_\nu^0$
results in ${\mathcal H}_1^0 {\mathcal H}_1^0 z_2^0$
which is invisible similarly to $A_\nu^0 \to \nu\overline{\nu}$.
}.
 If $y_\nu$ is large enough for $\mu \to e\gamma$
to be discovered in near future,
the process in Fig.~\ref{Fig:LHC} becomes very rare
because $H_\nu^0 \to \nu\overline{\nu}$ is the dominant channel.
 Next,
when the mixings between $Z_2$-odd particles are negligible,
${\mathcal H}_2^0$ can decay only into ${\mathcal H}_1^0 h^0$
via $\lambda_{s\Phi 1\eta}^{}$
because ${\mathcal H}_2^0 \to {\mathcal H}_1^0 H^0$
is kinematically forbidden
for the values in Eq.~\eqref{Eq:benchmark}.
 Thus,
even if $\lambda_{s\Phi 1\eta}^{}$ is rather small,
the branching ratio for ${\mathcal H}_2^0 \to {\mathcal H}_1^0 h^0$
can be almost 100\,\%.
 As a result,
the process in Fig.~\ref{Fig:LHC}
can be free from the one-loop suppression
and smallness of coupling constants
($y_\nu$, $\lambda_{s\Phi 1\eta}^{}$, and $\lambda_{s\Phi 2\eta}^{}$)
which are used to suppress $v_\nu^{}$.
 The cross section of
$pp \to H_\nu^+ H_\nu^0 + H_\nu^- H_\nu^0$
for the masses in Eq.~\eqref{Eq:benchmark}
is $7\,\text{fb}$ at the LHC with $\sqrt{s} = 14\,\TeV$.
 The SM background events come from
$t\overline{t}$, $WZ$, and $t\overline{b}$.
 Cross sections for
$pp \to t\overline{t}$, $W^+Z + W^-Z$, and $t\overline{b}+\overline{t} b$
at the LHC with $\sqrt{s} = 14\,\TeV$
are $833\,\text{pb}$~\cite{Bonciani:1998vc},
$55.4\,\text{pb}$~\cite{Campbell:1999ah},
and $3.91\,\text{pb}$~\cite{Kidonakis:2010tc},
respectively.
 Detailed analysis on kinematic cuts of the background events
is beyond the scope of this paper.

 If Nature chooses a parameter set
for which the process in Fig.~\ref{Fig:LHC} is not possible,
the deviation from the \nTHDM\
would be the increase of new scalar particles
which might be discovered directly
and/or change predictions in the \nTHDM\
about e.g.\ $h^0 \to \gamma\gamma$.

\section{Conclusions and Discussion}
\label{Sec:concl}

 The \nTHDM\ is a new physics model
where masses of Dirac neutrinos are generated
by a VEV~($v_\nu^{}$) of the second $\SU(2)_L$-doublet scalar field $\Phi_\nu$
which has a Yukawa interaction with only $\nu_R$
because of a global $\U(1)_X$ symmetry in the Lagrangian.
 We have presented a simple extension of the \nTHDM\
by introducing the third $\SU(2)_L$-doublet scalar field $\eta$
and two neutral $\SU(2)_L$ singlet fields ($s_1^0$ and $s_2^0$).
 Although the global $\U(1)_X$ is broken by a VEV of $s_1^0$,
there remains a residual $Z_2$ symmetry
under which $\eta$ and $s_2^0$ are $Z_2$-odd particles.
 These $Z_2$-odd particles provide a dark matter candidate.
 The $v_\nu^{}$ for neutrino masses can be suppressed
without requiring very heavy particles
because the VEV is generated at the one-loop level.

 A possible signature of the deviation from the \nTHDM\ at the LHC
is $\overline{\ell} j_b j_b \cancel{E}_T$ via
$pp \to H_\nu^+ H_\nu^0$ followed by
$H_\nu^+ \to \overline{\ell} \nu$
and
$H_\nu^0 \to {\mathcal H}_1^0 {\mathcal H}_2^0
\to {\mathcal H}_1^0 {\mathcal H}_1^0 h^0
\to {\mathcal H}_1^0 {\mathcal H}_1^0 b \overline{b}$.
 Coupling constants
which control $H_\nu^0 \to {\mathcal H}_1^0 {\mathcal H}_2^0$
and ${\mathcal H}_2^0 \to {\mathcal H}_1^0 h^0$
are the ones used in the one-loop diagram
which is the key to generate $v_\nu$.

\begin{acknowledgments}
 We thank Koji Tsumura and Natsumi Nagata for usefull comments.
 The work of S.K.\ was supported by
Grant-in-Aid for Scientific Research 
Nos.~22244031, 23104006, and 24340036.
 The work of H.S.\ was supported
by Grant-in-Aid for Young Scientists~(B)
No.~23740210.
\end{acknowledgments}

\appendix

\section*{Appendix}

\subsection{Scalar Potential}

 The scalar potential $V$ is given by
\begin{eqnarray}
V
&=&
 V_2 + V_3 + V_4 ,\\
V_2
&=&
 - \mu_{s1}^2 | s_1^0 |^2
 + \mu_{s2}^2 | s_2^0 |^2
 - \mu_{\Phi 1}^2 \Phi^\dagger \Phi
 + \mu_{\Phi 2}^2 \Phi_\nu^\dagger \Phi_\nu
 + \mu_\eta^2 \eta^\dagger \eta ,\\
V_3
&=&
 - \mu\, s_1^{0\ast} (s_2^0)^2 + \text{h.c.} ,\\
V_4
&=&
 \lambda_{s1} | s_1^0 |^4
 + \lambda_{s2} | s_2^0 |^4
 + \lambda_{s12} | s_1^0 |^2 | s_2^0 |^2
\nonumber\\
&&{}
 + \lambda_{\Phi 1} (\Phi^\dagger \Phi)^2
 + \lambda_{\Phi 2} (\Phi_\nu^\dagger \Phi_\nu)^2
 + \lambda_\eta (\eta^\dagger \eta)^2
\nonumber\\
&&{}
 + \lambda_{\Phi 12} (\Phi^\dagger \Phi) (\Phi_\nu^\dagger \Phi_\nu)
 + \lambda_{\Phi 1 \eta} (\Phi^\dagger \Phi) (\eta^\dagger \eta)
 + \lambda_{\Phi 2 \eta} (\Phi_\nu^\dagger \Phi_\nu) (\eta^\dagger \eta)
\nonumber\\
&&{}
 + \lambda_{\Phi 12}^\prime (\Phi^\dagger \Phi_\nu) (\Phi_\nu^\dagger \Phi)
 + \lambda_{\Phi 1 \eta}^\prime (\Phi^\dagger \eta) (\eta^\dagger \Phi)
 + \lambda_{\Phi 2 \eta}^\prime (\Phi_\nu^\dagger \eta) (\eta^\dagger \Phi_\nu)
\nonumber\\
&&{}
 + \left(
    \lambda_{\Phi 12\eta} (\Phi_\nu^\dagger \eta) (\Phi^\dagger \eta)
    + \text{h.c.}
   \right)
\nonumber\\
&&{}
 + \lambda_{s1 \Phi 1} | s_1^0 |^2 (\Phi^\dagger \Phi)
 + \lambda_{s1 \Phi 2} | s_1^0 |^2 (\Phi_\nu^\dagger \Phi_\nu)
 + \lambda_{s1 \eta} | s_1^0 |^2 (\eta^\dagger \eta)
\nonumber\\
&&{}
 + \lambda_{s2 \Phi 1}| s_2^0 |^2 (\Phi^\dagger \Phi)
 + \lambda_{s2 \Phi 2} | s_2^0 |^2 (\Phi_\nu^\dagger \Phi_\nu)
 + \lambda_{s2 \eta} | s_2^0 |^2 (\eta^\dagger \eta)
\nonumber\\
&&{}
 + \left(
    \lambda_{s \Phi1 \eta}\,
    s_1^{0\ast} (s_2^0)^\ast \Phi^\dagger \eta
    + \text{h.c.}
   \right)
 + \left(
    \lambda_{s \Phi2 \eta}\,
    s_1^0 s_2^0 \Phi_\nu^\dagger \eta
    + \text{h.c.}
   \right) .
\end{eqnarray}
 Actually,
the following simplified $V_4$
is sufficient for our discussion:
\begin{eqnarray}
V_4(\text{simplified})
&=&
 \lambda_{\Phi 1} (\Phi^\dagger \Phi)^2
 + \lambda_{s2} | s_2^0 |^4
 + \lambda_{s2\Phi 1}^{} | s_2^0 |^2 (\Phi^\dagger \Phi)
\nonumber\\
&&{}
 + \lambda_\eta (\eta^\dagger \eta)^2
 + \lambda_{\Phi 1\eta} (\Phi^\dagger \Phi) (\eta^\dagger \eta)
 + \lambda_{\Phi 2\eta} (\Phi^\dagger \eta) (\eta^\dagger \Phi)
\nonumber\\
&&{}
 + \lambda_{s1} | s_1^0 |^4
 + \lambda_{\Phi 2} (\Phi_\nu^\dagger \Phi_\nu)^2
\nonumber\\
&&{}
 + \left(
    \lambda_{s \Phi1 \eta}\,
    s_1^{0\ast} (s_2^0)^\ast \Phi^\dagger \eta
    + \text{h.c.}
   \right)
 + \left(
    \lambda_{s \Phi2 \eta}\,
    s_1^0 s_2^0 \Phi_\nu^\dagger \eta
    + \text{h.c.}
   \right) .
\end{eqnarray}

\subsection{Masses of Scalar Bosons}

 Scalar fields are decomposed as follows:
$\phi^0
=
 \frac{1}{\sqrt{2}}
 \left(
  v + \phi_r^0 + i \phi_i^0
 \right)
$ ,
$\phi_\nu^0
=
 \frac{1}{\sqrt{2}}
 \left(
  v_\nu + \phi_{\nu r}^0 + i \phi_{\nu i}^0
 \right)
$,
$s_1^0
=
 \frac{1}{\sqrt{2}}
 \left(
  v_s + s_{1r}^0 + i s_{1i}^0
 \right)
$,
$\eta^0
=
 \frac{1}{\sqrt{2}}
 \left(
  \eta_r^0 + i \eta_i^0
 \right)
$,
$s_2^0
=
 \frac{1}{\sqrt{2}}
 \left(
  s_{2r}^0 + i s_{2i}^0
 \right)
$.
 We ignore $v_\nu$ in the following formulae.

 The mass matrix for $(s_{2r}^0 , \eta_r^0)$
is obtained as
\begin{eqnarray}
M_{\mathcal H}^2
=
 \begin{pmatrix}
  m_{s2}^2 - \sqrt{2}\, \mu v_s \
  &
   \displaystyle
   \frac{1}{\,2\,} \lambda_{s\Phi 1 \eta}\, v\, v_s^{}\\[1mm]
  \displaystyle
  \frac{1}{\,2\,} \lambda_{s\Phi 1 \eta}\, v\, v_s^{}
  &
   m_\eta^2
 \end{pmatrix} ,
\end{eqnarray}
where
$m_{s2}^2 \equiv
 \mu_{s2}^2
 + \frac{1}{\,2\,}
   \left(
    \lambda_{s2\Phi 1} v^2
    + \lambda_{s12} v_s^2
   \right)$
and
$m_\eta^2
\equiv
 \mu_\eta^2
 + \frac{1}{\,2\,}
   \Bigl\{
    \left(
     \lambda_{\Phi 1\eta} + \lambda_{\Phi 1\eta}^\prime
    \right) v^2
    + \lambda_{s1\eta} v_s^2
   \Bigr\}$.
 On the other hand,
The mass matrix for $(s_{2i}^0 , \eta_i^0)$
results in
\begin{eqnarray}
M_{\mathcal A}^2
=
 \begin{pmatrix}
  m_{s2}^2 + \sqrt{2}\, \mu v_s \
  &
   \displaystyle
   \frac{1}{\,2\,} \lambda_{s\Phi 1 \eta}\, v\, v_s^{}\\[1mm]
  \displaystyle
  \frac{1}{\,2\,} \lambda_{s\Phi 1 \eta}\, v\, v_s^{}
  &
   m_\eta^2
 \end{pmatrix}\!\! .
\end{eqnarray}
 Notice that the difference between
$M_{\mathcal H}^2$ and $M_{\mathcal A}^2$
exists only in the $(1,1)$ element
as $(M_{\mathcal A}^2)_{11} = (M_{\mathcal H}^2)_{11} + 2\sqrt{2}\,\mu v_s^{}$.
 Mass eigenstates (${\mathcal H}_1^0$ and ${\mathcal H}_2^0$)
of $Z_2$-odd CP-even scalar bosons are given by
\begin{eqnarray}
\begin{pmatrix}
{\mathcal H}_1^0\\[1mm]
{\mathcal H}_2^0
\end{pmatrix}
=
 \begin{pmatrix}
  \cos\theta_0^\prime & -\sin\theta_0^\prime\\
  \sin\theta_0^\prime & \cos\theta_0^\prime
 \end{pmatrix}
 \begin{pmatrix}
  s_{2r}^0\\[1mm]
  \eta_r^0
 \end{pmatrix} , \quad
\tan(2\theta_0^\prime)
= \frac{ \lambda_{s\Phi 1 \eta}\, v\, v_s^{} }
       { m_\eta^2 - m_{s2}^2 + \sqrt{2}\, \mu v_s^{} } ,
\end{eqnarray}
while
mass eigenstates (${\mathcal H}_1^0$ and ${\mathcal H}_2^0$)
of $Z_2$-odd CP-odd scalar bosons are obtained as
\begin{eqnarray}
\begin{pmatrix}
{\mathcal A}_1^0\\[1mm]
{\mathcal A}_2^0
\end{pmatrix}
=
 \begin{pmatrix}
  \cos\theta_A^\prime & -\sin\theta_A^\prime\\
  \sin\theta_A^\prime & \cos\theta_A^\prime
 \end{pmatrix}
 \begin{pmatrix}
  s_{2i}^0\\[1mm]
  \eta_i^0
 \end{pmatrix} , \quad
\tan(2\theta_A^\prime)
= \frac{ \lambda_{s\Phi 1 \eta}\, v\, v_s^{} }
       { m_\eta^2 - m_{s2}^2 - \sqrt{2}\, \mu v_s^{} } .
\end{eqnarray}
 The mass eigenstate ${\mathcal H}^\pm$
of $Z_2$-odd charged scalar boson is identical to $\eta^\pm$:
\begin{eqnarray}
 {\mathcal H}^\pm = \eta^\pm .
\end{eqnarray}
 Masses of these $Z_2$-odd scalar bosons are calculated as
\begin{eqnarray}
m_{{\mathcal H}_1^0}^2
&=&
 \frac{1}{\,2\,}
 \biggl\{
  m_\eta^2 + m_{s2}^2 - \sqrt{2}\, \mu v_s^{}
  - \sqrt{
          \bigl(
           m_\eta^2 - m_{s2}^2 + \sqrt{2}\, \mu v_s^{}
          \bigr)^2
          + \lambda_{s\Phi 1 \eta}^2\, v^2 v_s^2
         }\,
 \biggr\} ,
\label{Eq:mH1}\\
m_{{\mathcal H}_2^0}^2
&=&
 \frac{1}{\,2\,}
 \biggl\{
  m_\eta^2 + m_{s2}^2 - \sqrt{2}\, \mu v_s^{}
  + \sqrt{
          \bigl(
           m_\eta^2 - m_{s2}^2 + \sqrt{2}\, \mu v_s^{}
          \bigr)^2
          + \lambda_{s\Phi 1 \eta}^2\, v^2 v_s^2
         }\,
 \biggr\} ,
\label{Eq:mH2}\\
m_{{\mathcal A}_1^0}^2
&=&
 \frac{1}{\,2\,}
 \biggl\{
  m_\eta^2 + m_{s2}^2 + \sqrt{2}\, \mu v_s^{}
  - \sqrt{
          \bigl(
           m_\eta^2 - m_{s2}^2 - \sqrt{2}\, \mu v_s^{}
          \bigr)^2
          + \lambda_{s\Phi 1 \eta}^2\, v^2 v_s^2
         }\,
 \biggr\} ,
\label{Eq:mA1}\\
m_{{\mathcal A}_2^0}^2
&=&
 \frac{1}{\,2\,}
 \biggl\{
  m_\eta^2 + m_{s2}^2 + \sqrt{2}\, \mu v_s^{}
  + \sqrt{
          \bigl(
           m_\eta^2 - m_{s2}^2 - \sqrt{2}\, \mu v_s^{}
          \bigr)^2
          + \lambda_{s\Phi 1 \eta}^2\, v^2 v_s^2
         }\,
 \biggr\} ,
\label{Eq:mA2}\\
m_{{\mathcal H}^\pm}^2
&=&
 m_\eta^2
 - \frac{1}{\,2\,} \lambda_{\Phi 1 \eta}^\prime v^2 .
\end{eqnarray}

 Next,
the mass matrix for $(\phi_r^0 , s_{1r}^0)$
is given by
\begin{eqnarray}
M_H^2
=
 \begin{pmatrix}
  2 \lambda_{\Phi 1} v^2
  & \lambda_{s1\Phi 1}\, v\, v_s\\
  \lambda_{s1\Phi 1}\, v\, v_s
  & 2 \lambda_{s1} v_s^2
 \end{pmatrix} .
\end{eqnarray}
 Notice that
$\phi_{\nu r}^0$ does not mix with them
when we ignore $v_\nu$.
 Mass eigenstates ($h^0$, $H^0$, and $H_\nu^0$)
of $Z_2$-even CP-even scalar bosons are given by
\begin{eqnarray}
\begin{pmatrix}
h^0\\[1mm]
H^0
\end{pmatrix}
&=&
 \begin{pmatrix}
  \cos\theta_0 & -\sin\theta_0\\
  \sin\theta_0 & \cos\theta_0
 \end{pmatrix}
 \begin{pmatrix}
  \phi_r^0\\[1mm]
  s_{1r}^0
 \end{pmatrix} , \quad
\tan(2\theta_0)
= \frac{ \lambda_{s1 \Phi 1}\, v\, v_s^{} }
       {
        \lambda_{s1} v_s^2 - \lambda_{\Phi 1} v^2
       } ,\\
H_\nu^0
&=&
 \phi_{\nu r}^0 .
\end{eqnarray}
 The Nambu-Goldstone boson $z_2^0$ for the $\U(1)_X$ breaking,
a $Z_2$-even CP-odd scalar boson $A_\nu^0$,
and the $Z_2$-even charged scalar boson $H_\nu^\pm$
are defined as follows:
\begin{eqnarray}
z_2^0 = s_{1i}^0 , \quad
A_\nu^0 = \phi_{\nu i}^0 , \quad
H_\nu^\pm = \phi_\nu^\pm .
\end{eqnarray}
 Masses of these $Z_2$-even scalar bosons are
calculated as
\begin{eqnarray}
m_{h^0}^2
&=&
 \lambda_{s1} v_s^2 + \lambda_{\Phi 1} v^2
 - \sqrt{
         \left\{
          \lambda_{s1} v_s^2 - \lambda_{\Phi 1} v^2
         \right\}^2
         + \lambda_{s1 \Phi 1}^2\, v^2\, v_s^2
        } ,\\
m_{H^0}^2
&=&
 \lambda_{s1} v_s^2 + \lambda_{\Phi 1} v^2
 + \sqrt{
         \left\{
          \lambda_{s1} v_s^2 - \lambda_{\Phi 1} v^2
         \right\}^2
         + \lambda_{s1 \Phi 1}^2\, v^2\, v_s^2
        } , \\
m_{z_2^0}^2
&=&
 0 ,\\
m_{H_\nu^0}^2
=
m_{A_\nu^0}^2
&=&
 \mu_{\Phi 2}^2
 + \frac{1}{\,2\,}
   \Bigl\{
    (\lambda_{\Phi 12} + \lambda_{\Phi 12}^\prime) v^2
    + \lambda_{s1 \Phi 2} v_s^2
   \Bigr\} ,\\
m_{H_\nu^\pm}^2
&=&
 \mu_{\Phi 2}^2
 + \frac{1}{\,2\,}
   \left\{
    \lambda_{\Phi 12}^{} v^2
    + \lambda_{s1 \Phi2}^{} v_s^2
   \right\} .
\end{eqnarray}


\end{document}